Systematic Review

# A Systematic Review of Generative AI for Teaching and Learning Practice

**Bayode Ogunleye [1,*], Kudirat Ibilola Zakariyyah [1], Oluwaseun Ajao [2], Olakunle Olayinka [3] and Hemlata Sharma [4]**

[1] School of Architecture, Technology & Engineering, University of Brighton, Brighton BN2 4GJ, UK
[2] Department of Computing & Mathematics, Manchester Metropolitan University, Manchester M1 5GD, UK
[3] School of Computer Science, The University of Sheffield, Sheffield, S1 4DP, UK
[4] Department of Computing, Sheffield Hallam University, Sheffield S1 2NU, UK
* Correspondence: b.ogunleye@brighton.ac.uk

**Abstract:** The use of generative artificial intelligence (GenAI) in academia is a subjective and hotly debated topic. Currently, there are no agreed guidelines towards the usage of GenAI systems in higher education (HE) and, thus, it is still unclear how to make effective use of the technology for teaching and learning practice. This paper provides an overview of the current state of research on GenAI for teaching and learning in HE. To this end, this study conducted a systematic review of relevant studies indexed by Scopus, using the preferred reporting items for systematic reviews and meta-analyses (PRISMA) guidelines. The search criteria revealed a total of 625 research papers, of which 355 met the final inclusion criteria. The findings from the review showed the current state and the future trends in documents, citations, document sources/authors, keywords, and co-authorship. The research gaps identified suggest that while some authors have looked at understanding the detection of AI-generated text, it may be beneficial to understand how GenAI can be incorporated into supporting the educational curriculum for assessments, teaching, and learning delivery. Furthermore, there is a need for additional interdisciplinary, multidimensional studies in HE through collaboration. This will strengthen the awareness and understanding of students, tutors, and other stakeholders, which will be instrumental in formulating guidelines, frameworks, and policies for GenAI usage.

**Keywords:** artificial intelligence; generative AI; higher education; PRISMA; systematic literature review; teaching and learning; topic modelling

## 1. Introduction

Generative artificial intelligence (GenAI) tools have taken the world by storm, most especially, ChatGPT and now, Gemini [1]. The advancement in technology has raised strands of concern in various sectors, specifically, on the assumption that technology will replace peoples' jobs. A perceived predominant sector related to such effect is higher education [2]. The higher education (HE) sector contributes to every nation's economy, offering a wealth of benefits to society. HE plays a key role in enhancing social mobility, bolstering social capital, fostering political stability, reducing crime rates, promoting social unity, spurring innovation, and cultivating trust and tolerance [3]. The development of such technology can be traced back to the advent of large language models (LLMs) in 2018, when BERT was released [4]. Since then, several LLMs have been released including GPT [5]. GenAI tools rely on these LLMs to perform the task they are developed for. For example, ChatGPT relies on the GPT series to perform its task. LLMs are trained on a large number of parameters (data), including text and images [5–7]. By processing a huge amount of data, LLMs learn the statistical relationships, patterns, and structure within datasets, which enables them to predict or generate relevant and meaningful content in response to user requests. Thus, they are capable of performing various complex tasks [6–9].

However, concerns including hallucinations [10], bias [11], ethical and privacy concerns [12,13], accidental plagiarism [14], and academic integrity [15–20] have been raised regarding GenAI tools; such tools have been praised for their potential benefits in relation to HE. For example, Daun et al. [2] demonstrated the use of GenAI for teaching and learning within the context of software engineering education. They showed that GenAI tools like ChatGPT can be used to find literature, answer student questions, support code implementation, and generate exercises. Kurtz et al. [21] synthesised the literature and found out that GenAI offers opportunities to enhance students' learning experiences by facilitating learning environments tailored to students' educational needs. Their findings further suggest that the potential use of GenAI for student performance prediction offers an opportunity for early interventions, potentially reducing student churn and dropout rates. Atlas [22] reported the current application of GenAI in HE as automated essay scoring; personalised tutoring; research assistance; language translation; helping professors in creating their syllabus, quizzes, and exams; generating reports; and email and chatbot assistance. Pesovski et al. [23] added that GenAI provides an opportunity for affordable and sustainable personalised learning. Other notable benefits of GenAI in HE are creative writing and brainstorming [24], support for personalised tutoring [13], support for pro-gramming code development [25], essay grading [26], and it is useful for designing science units and rubrics [27]. In terms of user perceptions, Rajabi et al. [28] investigated both students' and teachers' perspectives on the integration of GenAI for teaching and learning within a post-secondary school environment using a qualitative dataset. The participants recognised the tool as an advanced search engine and emphasised that students are likely to use GenAI tools, irrespective of whether the tools are incorporated into HE courses or not. Their results showed that students and teachers have mixed perceptions about ChatGPT's usage in a post-secondary school setting. The findings by Lozano and Blanco-Fontao [29] showed that students have a positive perception of the utilisation of GenAI tools in HE. Most importantly, their findings suggest that ChatGPT is not perceived as a threat, causing the deterioration of the educational system. Moreover, Sánchez-Ruiz et al. [30] surveyed 110 students to gather students' perspectives on the impact and usage of ChatGPT. Their results showed that students have a positive opinion of ChatGPT. However, there are concerns about gaining problem-solving and creative skills.

Based on the challenges and promise GenAI offers to HE sectors, it is worth highlighting and synthesising the literature to understand the potential use, impact, and ethical issues posed by AI tools in the context of teaching and learning. To this end, this paper aims to conduct a systematic literature review on the use of GenAI tools in HE, provide an overview of the current state of research on GenAI for teaching and learning, and offer insights into future research directions. By doing so, this study formulates two research questions (RQ) to be answered, as follows:

- RQ1. What is the evolutionary productivity in the field in terms of the most influential journals, most cited articles, and authors, including geographical distribution of authorship?
- RQ2. What are the main trends and core themes emerging from the extant literature?

To the best of the researchers' knowledge, only limited studies were found to have conducted a systematic review of the literature on GenAI in education. The authors in [31] conducted a tertiary systematic review of AI tools in education. Sullivan et al. [32] used a systematic search to review English language newspapers and online news sources about how ChatGPT is disrupting HE across selected countries. Furthermore, Bahroun et al. [33] adopted the preferred reporting items for systematic reviews and meta-analyses (PRISMA) framework for the selection of literature on GenAI use in education. These authors reviewed a total of 207 research papers using bibliometric and content analysis to explore GenAI's transformative impact in specific educational domains, including medical education and engineering education. However, this study differs as it centres on a systematic review specifically on the use of GenAI for teaching and learning practice in HE. In addition, this paper adopts a topic modelling (TM) approach to distil information from the literature and report core themes. To conclude, relevant analysis of the data collected is performed and the main contributions of this paper are described, as follows:

- This review provides a comprehensive overview of the current state of research on GenAI for teaching and learning in HE, this helps researchers to identify the evolutionary progression (most influential journals, articles, authors, including geographical distribution of authorship), prevailing topics, and research directions within the field;
- This review synthesises the findings to generate insights into a holistic perspective on the potential, effectiveness, and limitations of GenAI use for teaching and learning in HE;
- This review identifies research gaps that require further investigation, guiding future research work.

## 2. Methodology

A systematic literature review on the use of GenAI in HE was conducted by adopting the PRISMA [34] guidelines. Scholarly articles (conference proceedings and journal papers) over the last 7 years were reviewed and analysed. As shown in Figure 1, the study made use of key search terms related to GenAI, teaching, and learning, and HE. The keywords, as shown in Table 1, were used to extract metadata from the relevant research papers (documents) in the Scopus database. The Scopus database was used due to its document volume, reliability, the accuracy of the information, and its advantage of using rigorous original metadata to associate people, published theories, and institutions [33,35]. The following subsections discuss the steps taken to achieve the data collection and analysis.

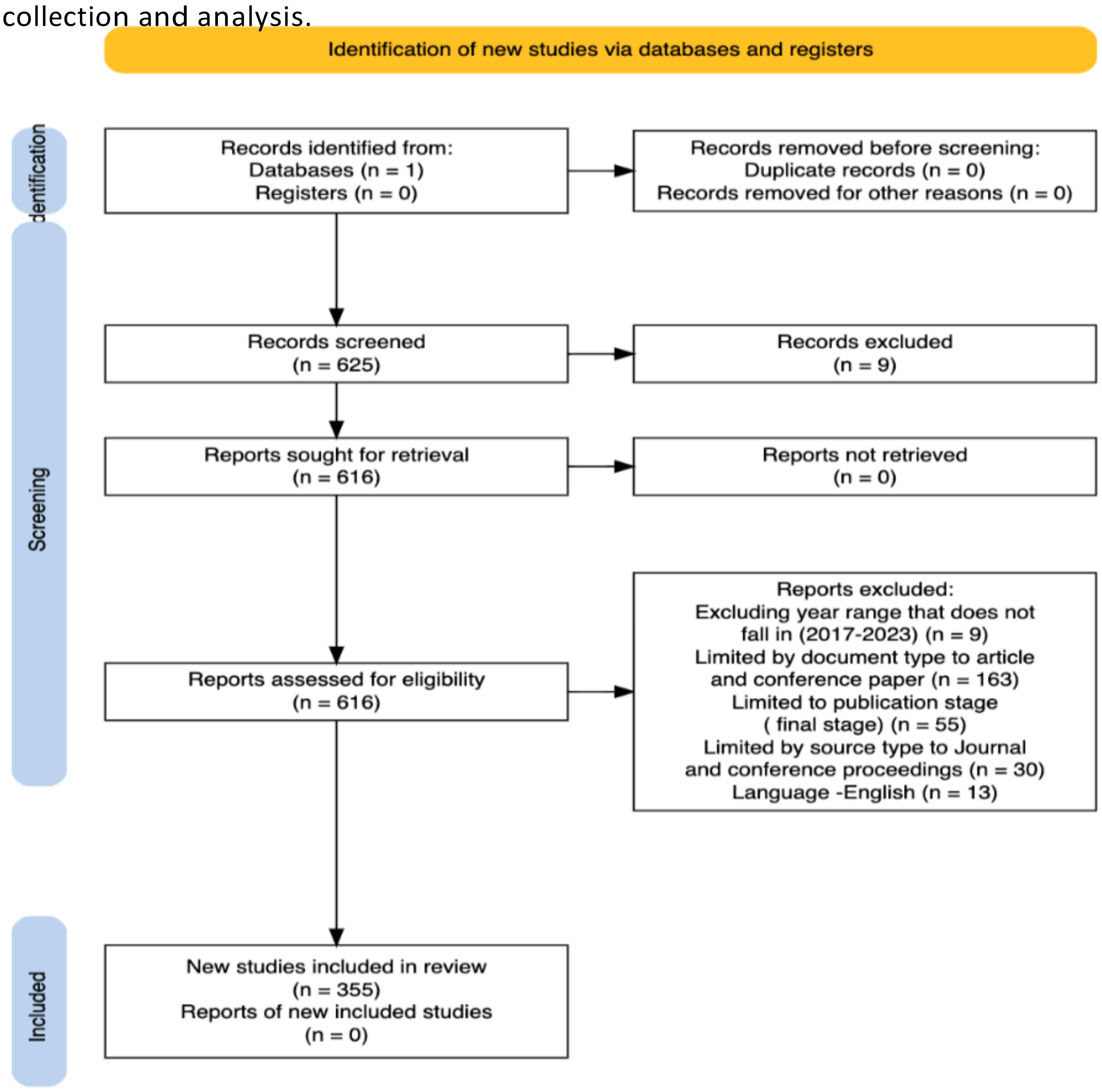

**Figure 1.** A systematic literature search using the PRISMA framework.

**Table 1.** Initial search strings.

| |
|---|
| (generative AND artificial AND intelligence OR generative AND ai OR genai OR gai) AND (assessment OR pedagogic OR student OR teaching AND learning OR teaching OR teacher) OR (llms OR language AND model) OR (academia OR education OR he OR higher AND education) |

### 2.1. Database Search and Eligibility Criteria

The search strategy adopted keywords such as "generative artificial intelligence", "assessment", "higher education", "HE", and "teaching and learning". Using the search expression in Table 1, the initial results generated 625 papers from the Scopus database. Next, the research papers were narrowed down to a period of 7 years (2017 to 2023). A period of 7 years was chosen because the concept of GenAI is relatively new, and although there are arguments that the wider the year range, the better the information that can be obtained from document convergence [36], from our review of the results, studies before 2017 were not relevant to the context of the study, thus 2017 was selected as the take-off point. Two of the authors reviewed the results generated independently against the inclusion and exclusion criteria and, in the case of disagreement, a third author helped to decide whether the paper met the criteria for inclusion. Using the inclusion and exclusion criteria presented in Table 2, a total of 355 papers were selected for analysis.

**Table 2.** Inclusion and exclusion criteria.

| Inclusion Criteria | Exclusion Criteria |
|---|---|
| Published between 2017 and 2023 | Published before 2017 |
| Publication should be peer reviewed | Not peer reviewed |
| Published in English | Papers not published in English due to authors' common language |
| Journal articles or conference papers | Editorials, meeting abstracts, workshop papers, posters, book reviews, and dissertations |

### 2.2. Data Quality Assessment

To ensure the study's reliability, Cohen's kappa inter-rater reliability assessment was conducted. The authors assessed the quality of the studies independently to ensure that the extracted papers were relevant. This was done by two researchers (BO and KZ). This process started by randomly selecting 20 documents from the data used for the analysis. These documents were then screened using the titles and abstracts. Having agreed on the criteria for inclusion or exclusion based on the titles and abstracts, the coding decisions of the two researchers (rater BO and rater KZ) were presented and assessed to determine the inter-rater reliability using Cohen's kappa (κ) value. In cases of disagreement, a third author helped to decide the outcome. Cohen's kappa coefficient depicts the value for the degree of consistency among the raters, that is the extent to which their measures are the same, based on the number of codes in the coding scheme and the value obtained [31]. For example, a kappa value of 0.40 to 0.60 is fair, 0.61 to 0.75 is good, while a value above 75 is excellent. After the necessary computation, the kappa value of 0.659 was arrived at. That is, the inter-rater reliability value is good for the coding of the inclusion and exclusion criteria, and there is consistency among the documents used for the analysis. This helped minimise the risk of bias and improved the data quality.

### 2.3. Bibliometric Approach

This study employed bibliometric indicators, such as the number of publications, number of citations, top cited documents (sources and authors), co-authorship (geographical distribution), and term co-occurrence (keywords, title, and abstract). This paper used suitable software, such as Python, Power BI, and VOS viewer, to present our findings, where appropriate.

### 2.4. Topic Modelling Approach

This study used a TM technique to distil the current state of research on GenAI in HE. The concept of TM is becoming popular for literature review analysis [37–40]. This is because the approach provides an automated and efficient way of uncovering hidden themes [6]. In this study, the researchers utilised a well-refined corpus fitted to the latent Dirichlet allocation (LDA) model to uncover latent themes from the research documents (abstract). LDA, proposed by Blei et al. [41], is a generative probabilistic model for topic extraction. The topic model captures the important intra-word/document statistical structure via a mixing distribution. LDA assumes words in each document are related and, thus, topic assignment strongly relies on local co-occurrence. The documents represent probability distributions over latent topics. While topics represent probability distributions over words. For the evaluation of our LDA topics, this paper employed the use of coherence scores, perplexity, and human interpretation. The coherence score indicates the data quality by comparing the semantic similarity between words in a topic. The coherence score measures how well the text aligns with human judgment and is closely correlated to human interpretation [6,42,43]. The coherence score is often interpreted as the higher the score, the better the model. This implies that a high score indicates that the grouped words are sensible, relevant, and consistent. Whilst a low coherence score means the topic is vague, noisy, or irrelevant [42]. This paper produced both a coherence score (cv) and a coherence score (Umass) to strengthen the model evaluation process. The coherence score (cv) calculates the probability of co-occurrence for word pairs generated using a sliding window and, thus, measures the mean cosine similarity between the word's feature vector and the topic's feature vector [43]. Whilst the coherence score (Umass) measures the word pair relationship based on document co-occurrence and, thus, for every K (number) topic, words are ordered (in descending order) based on the probability of a word for a given topic [43,44]. The perplexity indicates how well the model describes a document by computing the inverse log-likelihood of unseen data. Perplexity is often interpreted as the lower it is, the better the model. These metrics are considered appropriate for the performance evaluation of topic models [6].

## 3. Results and Discussion

The results of the analysis are presented in two subsections. The first (Section 3.1) provides insights into the bibliometric indicators. whilst the second (Section 3.2) presents the results from the TM.

### 3.1. Bibliometric Analysis Results

This section provides the results from the bibliometric analysis, as follows.

#### 3.1.1. Documents by Publication Type

Figure 2 shows that the extracted documents are 72% and 28% journal articles and conference proceedings, respectively. This implies that research on AI-related fields is evolving, and additional studies are needed.

#### 3.1.2. Publications per Year

From 2017 to 2023, there was a gradual increase in the number of papers published as both journal articles and conference papers over the years, with a significant jump in 2023, as shown in Figure 3. The year-on-year increase shows a growing interest in the area. However, in 2023, the significant jump in the number of publications from 38 to 273 can be explained by the launch of ChatGPT. In November 2022, OpenAI (the creators of the GPT series of LLMs) released ChatGPT to the public and, within two months of its release, it was estimated to have reached 100 million monthly active users [45]. This rapid adoption led academics to explore its impact on various aspects of teaching and learning.

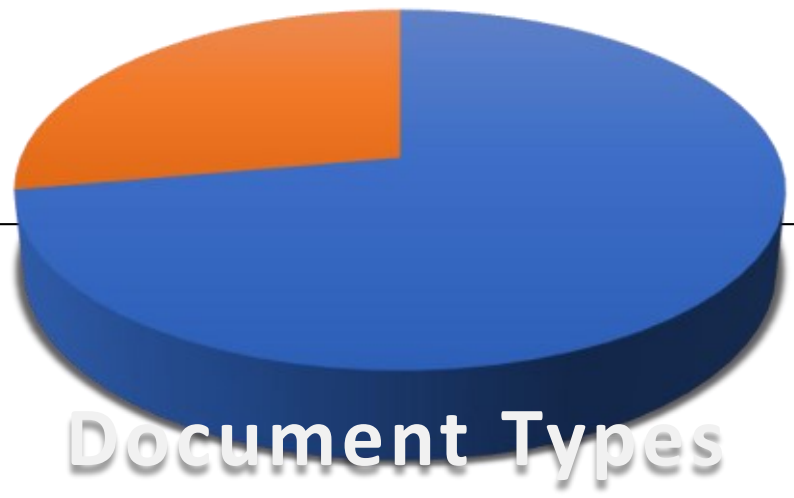

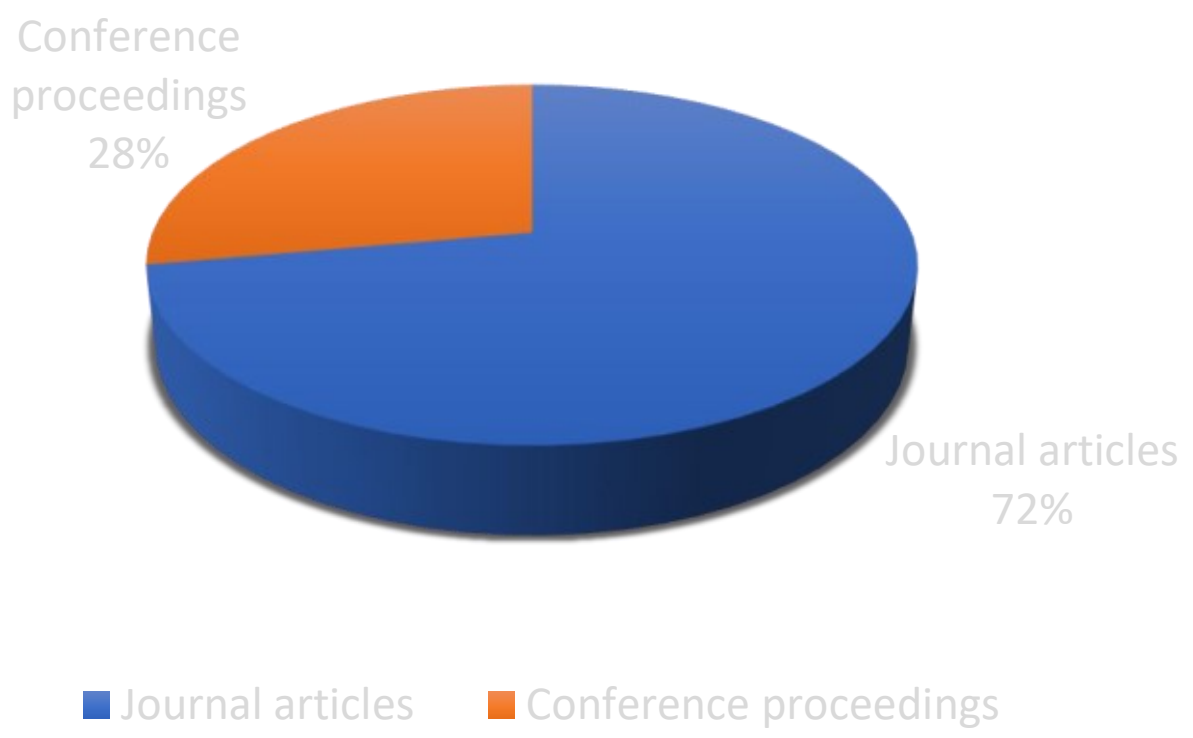

**Figure 2.** Documents by publication type.

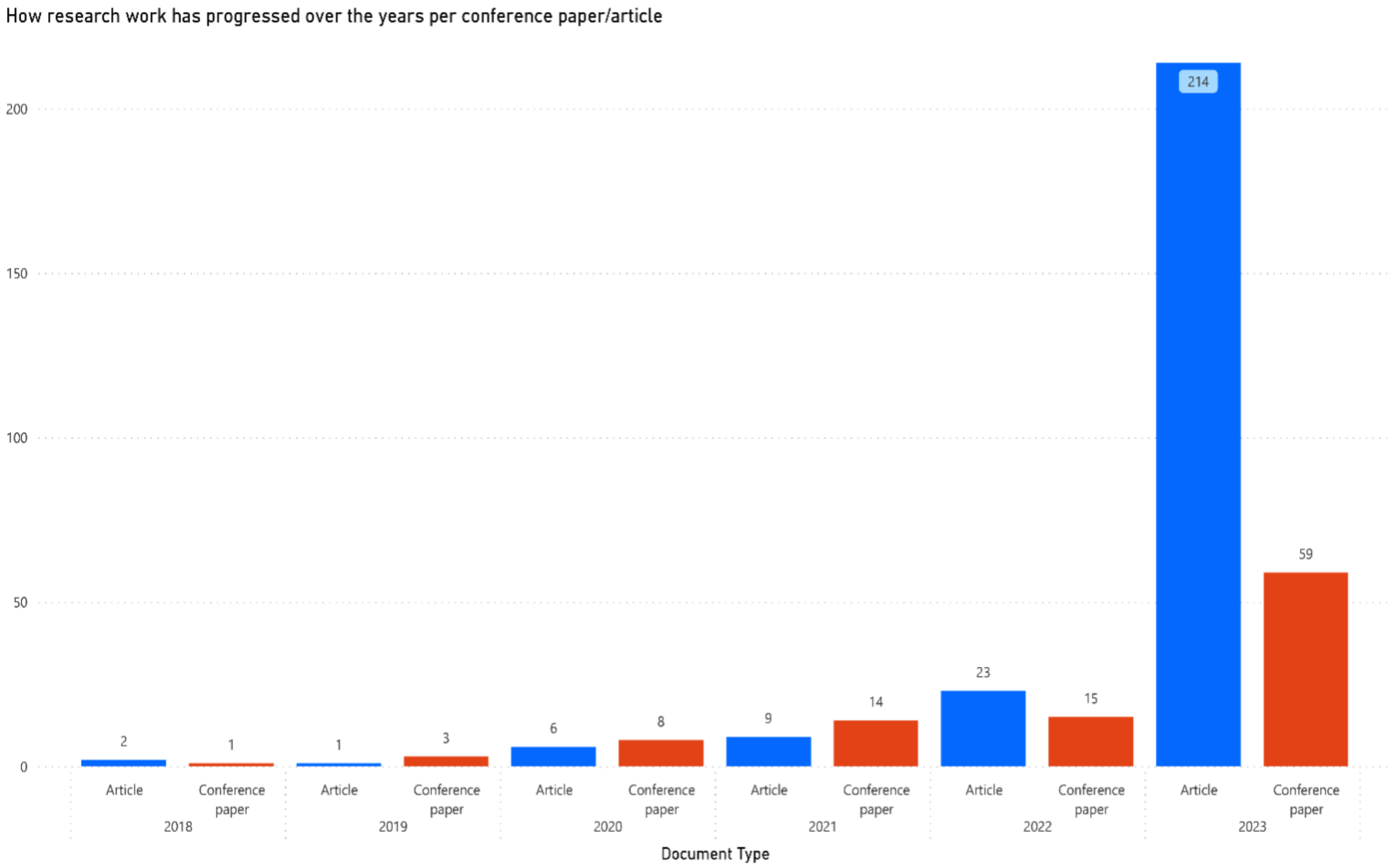

**Figure 3.** Publications per year.

3.1.3. Citation per Source Title

Figure 4 presents the top ten most cited sources. The most cited source is the Journal of Applied Learning and Teaching with 301 citations, and this is very closely followed by the International Journal of Information Management with 291 citations. The tenth most cited source is the IEEE Journal of Biomedical and Health Informatics, with 71 citations. It is interesting to note that there is currently an equal split between education and technology-focused journals in the ranked list. We posit that the adoption and impact of GenAI technologies in education for teaching, learning, and assessments will continue to grow.

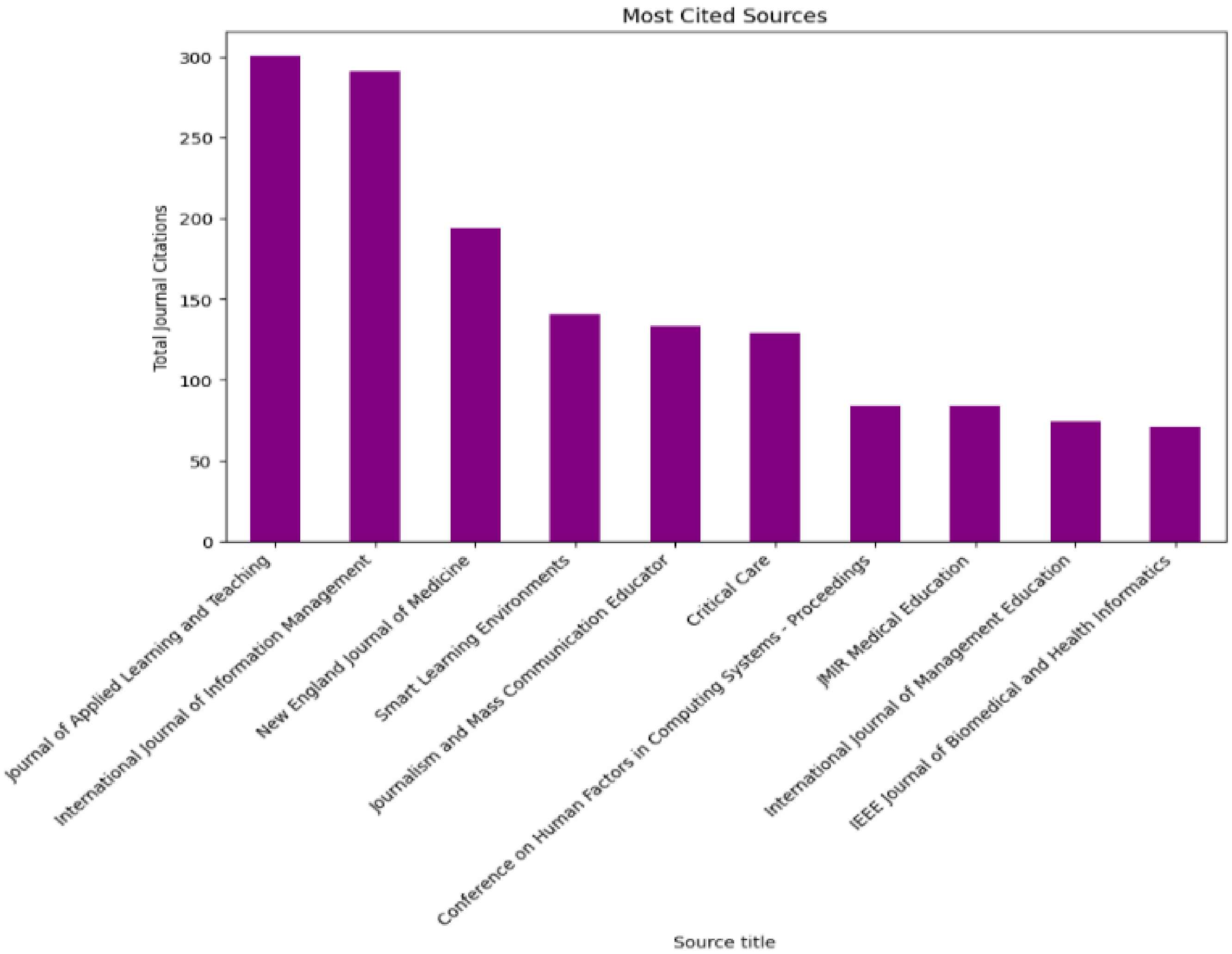

**Figure 4.** Top cited sources.

#### 3.1.4. Citations per First Authors

The analysis identified Yogesh K. Dwivedi as the first author with the highest number of citations at 291, followed by Jurgen Rudolph with 234 citations, as shown in Figure 5. As most of these citations have only been acquired since 2023, this shows that the conversations around GenAI in education are a trending topic amongst education researchers, technology experts, and industry practitioners.

#### 3.1.5. Publications/Citations per Year

The total number of documents used in this study is 355, with a total amount of citation of 2923, as shown in Table 3. Out of these documents, the chart in Figure 6 shows the distributions of the documents and the corresponding citations per year. From the chart, we observed that the documents in the year 2023 have the highest value at 273 (76.9% of the entire document). These 273 documents have 2078 citations (which is 71% of the total citations). Between 2018 and 2021, there was a steady progression of documents with citations received. However, in 2022, though the number of documents produced increased, the increase in the total number of citations was not commensurate. Nevertheless, the year 2023 ushered in extremely large documents with corresponding citations. This is not unusual, considering that GenAI, including ChatGPT, became popular in early 2023. Out of the 273 documents in the year 2023, Table 4 further depicts the 10 publications with the highest number of citations.

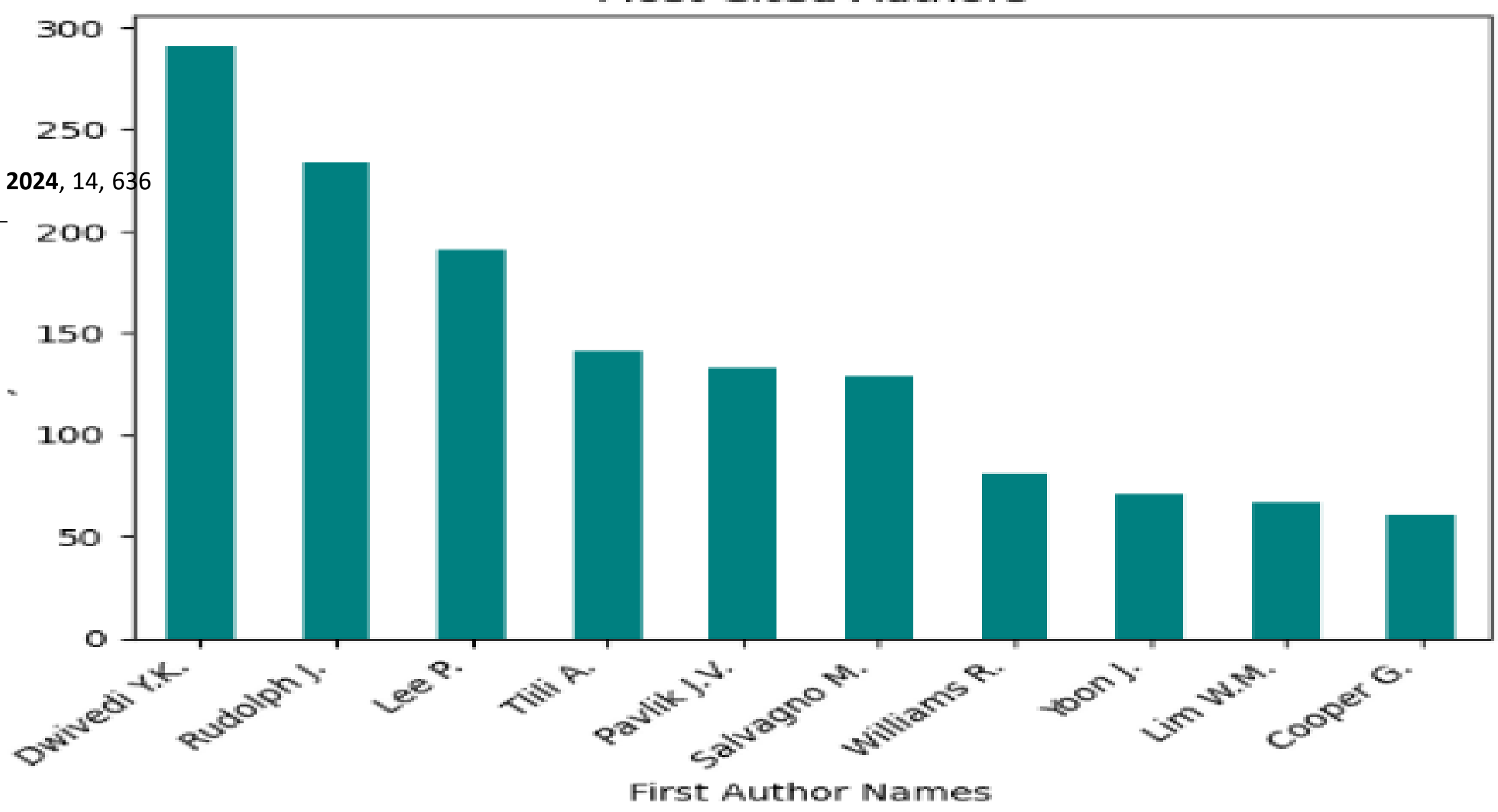

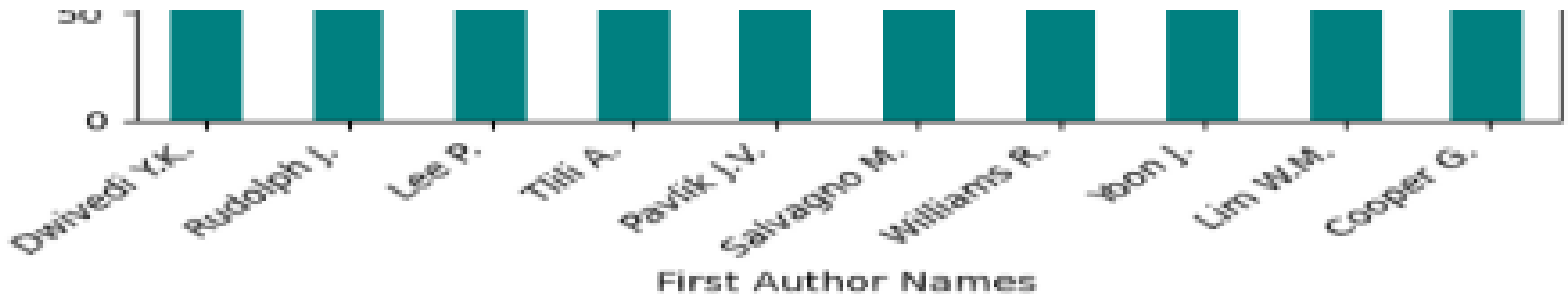

**Figure 5.** Top 10 most cited first authors.

**Table 3.** Total citations per year.

| Year | No. of Publications | Total Citations |
|---|---|---|
| 2018 | 3 | 49 |
| 2019 | 4 | 114 |
| 2020 | 14 | 196 |
| 2021 | 23 | 290 |
| 2022 | 38 | 196 |
| 2023 | 273 | 2078 |

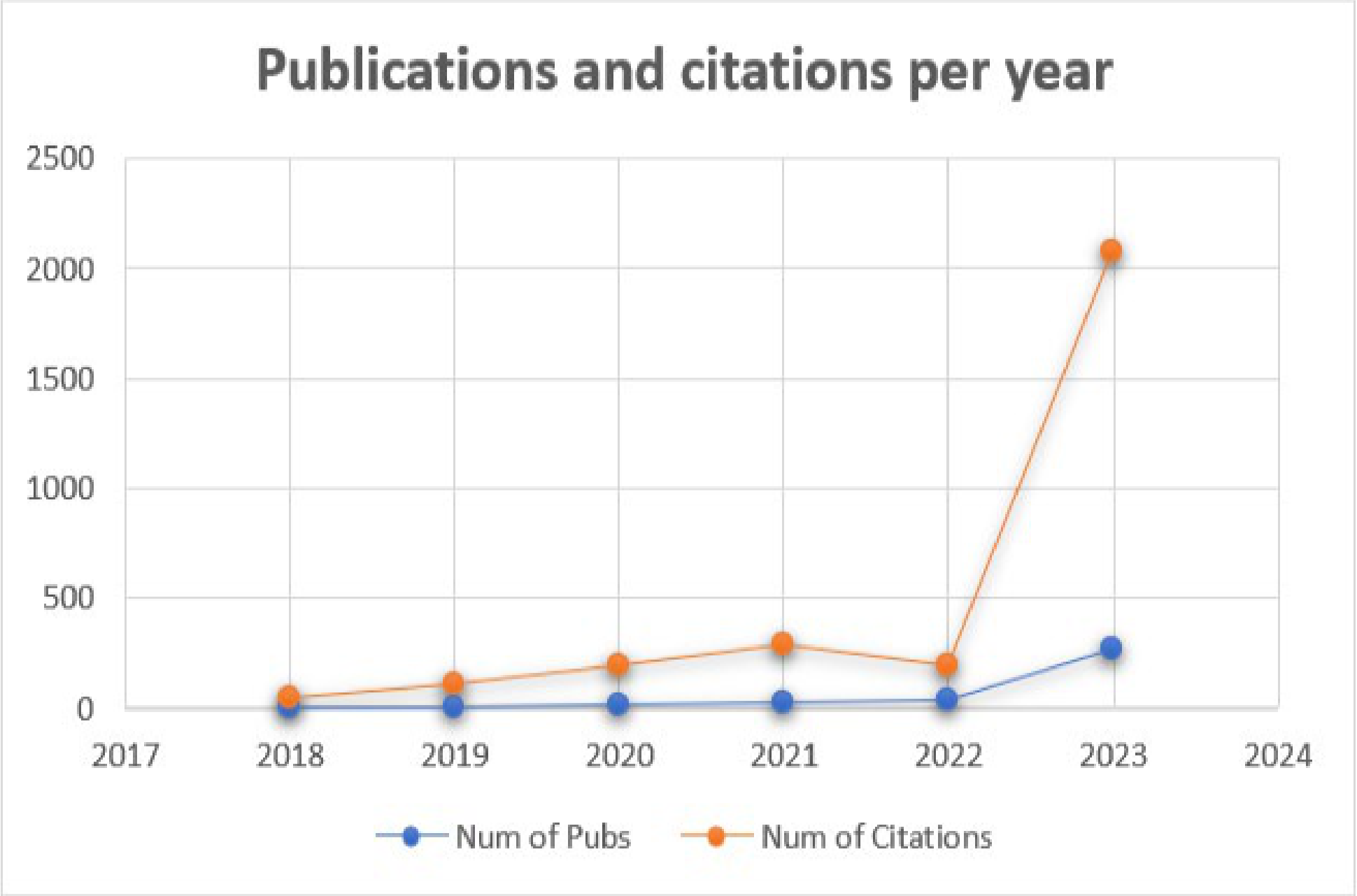

**Figure 6.** Publications and citations per year.

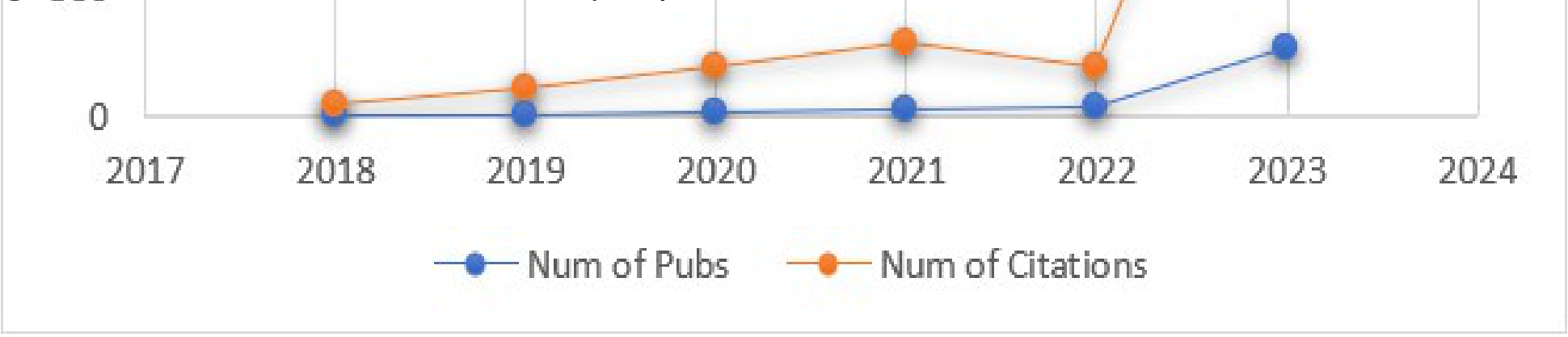

**Table 4.** Most cited publications (2023).

| Authors | Year | Title | Citations |
|---|---|---|---|
| Dwivedi et al. [46] | 2023 | "So what if ChatGPT wrote it?" Multidisciplinary perspectives on the opportunities, challenges, and implications of generative conversational AI for research, practice, and policy | 291 |
| Lee et al. [47] | 2023 | Benefits, limits, and risks of GPT-4 as an AI chatbot for medicine. | 191 |
| Rudolph et al. [48] | 2023 | ChatGPT: Bullshit spewer or the end of traditional assessments in higher education? | 152 |
| Tlili et al. [49] | 2023 | What if the devil is my guardian angel: ChatGPT as a case study of using chatbots in education | 141 |
| Pavlik [50] | 2023 | Collaborating With ChatGPT: Considering the implications of generative artificial intelligence for journalism and media education | 133 |
| Salvagno et al. [51] | 2023 | Can artificial intelligence help with scientific writing? | 129 |
| Rudolph et al. [52] | 2023 | War of the chatbots: Bard, Bing Chat, ChatGPT, Ernie, and beyond. The new AI gold rush and its impact on higher education | 82 |
| Lim et al. [53] | 2023 | Generative AI and the future of education: Ragnarök or reformation? A paradoxical perspective from management educators | 67 |
| Cooper [27] | 2023 | Examining science education in ChatGPT: An exploratory study of generative artificial intelligence | 61 |
| Crawford et al. [54] | 2023 | Leadership is needed for ethical ChatGPT: Character, assessment, and learning using artificial intelligence (AI) | 53 |

#### 3.1.6. Co-Authorship

This study produced a co-authorship visualisation map to understand co-authorship by countries. Figure 7 shows the geographical distribution according to co-authorship. For the analysis, the co-authorship country of origin was taken into consideration. Furthermore, a country is considered if it had at least three papers, which resulted in 35 countries being included for the analysis. As shown in Figure 7, the analysis evidenced that the most relevant countries in terms of authorship relationship, based on the number of papers (n) or citations, are the United States (n = 124, citations = 1598, total link strength = 88), Australia (n = 38, citations = 767, total link strength = 34), China (n = 36, citations = 249, total link strength = 32), and the United Kingdom (n = 32, citations = 462, total link strength = 28). The dominance of the United States in terms of the number of publications in this field is not surprising, as this is consistent with the findings of previous studies [55–60]. Overall, the volume of documents produced in the area of GenAI in HE sectors needs improvement, especially in the global south.

#### 3.1.7. Co-Occurrence

Figure 8 shows an overlay visualisation network map for the co-occurring keywords in the different years of publication. Firstly, this study used authors' keywords to investigate co-occurrence. The analysis produced five clusters and, as shown in Figure 8, artificial intelligence (n = 141) and ChatGPT (n = 126) are the highest co-occurring keywords. Specifically, in 2022, notable co-occurring keywords are grouped into clusters 2, 3, and 5, namely generative adversarial network, AI, chatbot, deep learning, machine learning, GPT-3, natural language processing, and language model. The author's keywords indicate publications related to the application of AI/machine learning to education. Whilst in 2023 (in yellow), the co-occurring keywords (grouped into clusters 1 and 4) are academic integrity, assessment, chatbots, ChatGPT, education, ethics, generative pre-trained transformer, GPT-4, higher education, medical education, OpenAI, and prompt engineering. This is not surprising due to the emergence of ChatGPT in November 2022, when this area attracted a lot of attention in terms of research and debate, especially in the medical education domain. To gain an in-depth understanding of keyword co-occurrence, this study performed title and abstract keyword co-occurrence analysis, as shown in Figure 9.

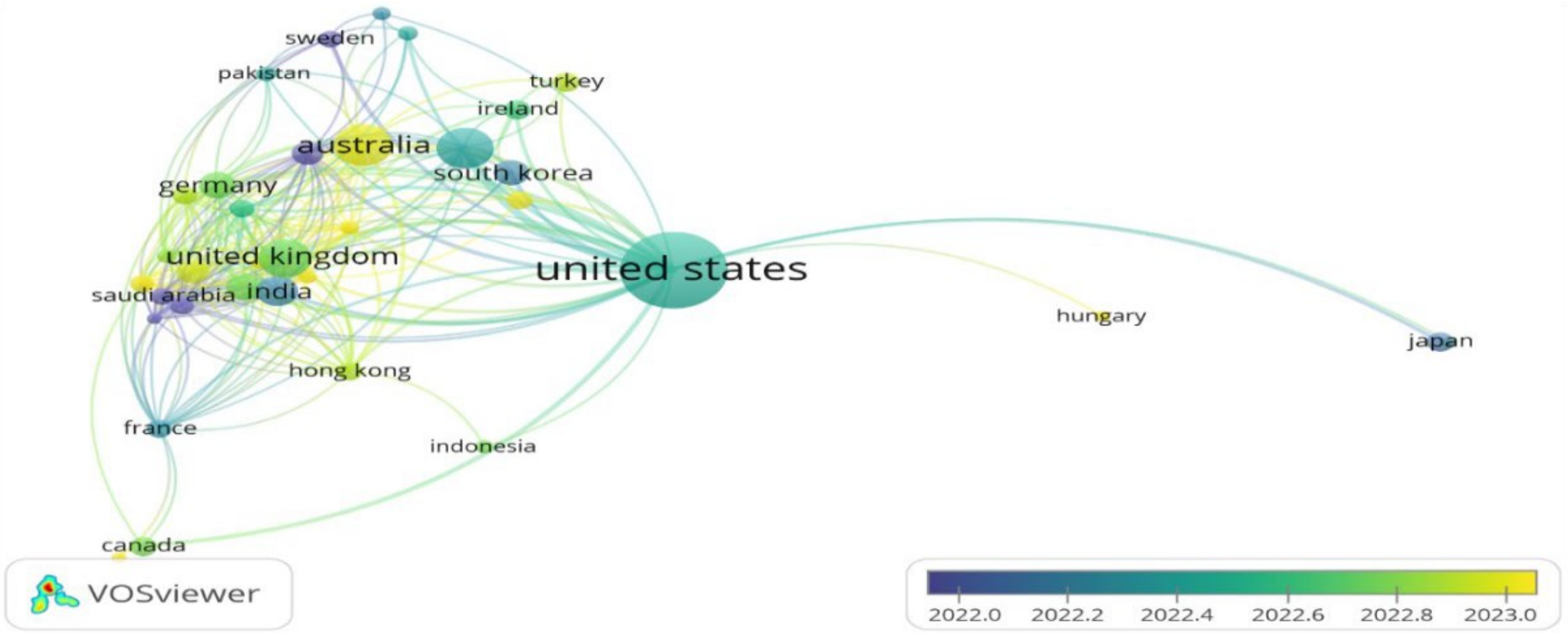

**Figure 7.** Co-authorship by countries.

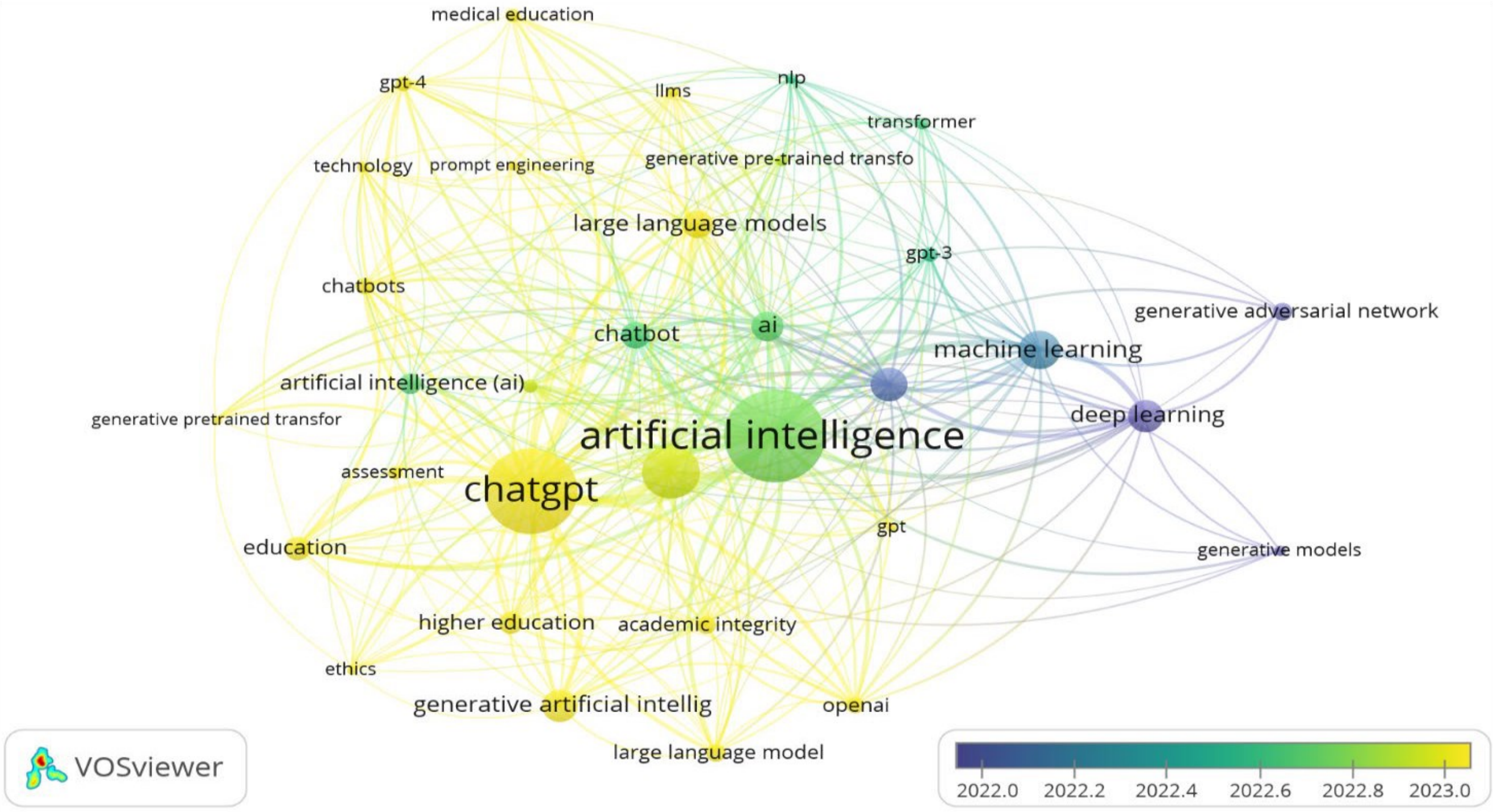

**Figure 8.** Map of co-occurrence (author keywords).

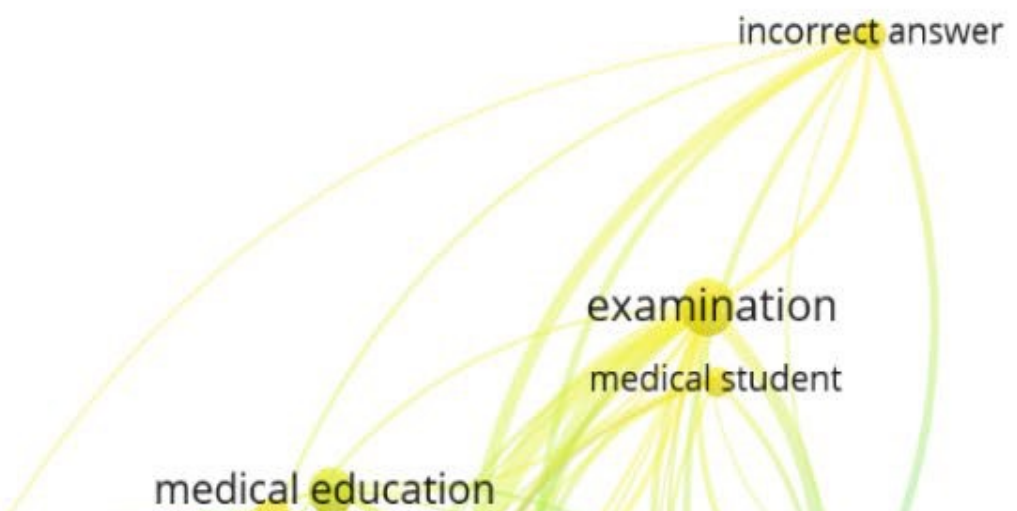

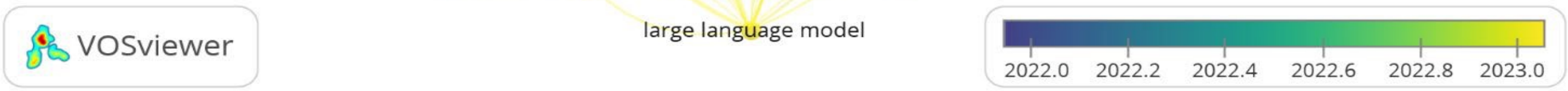

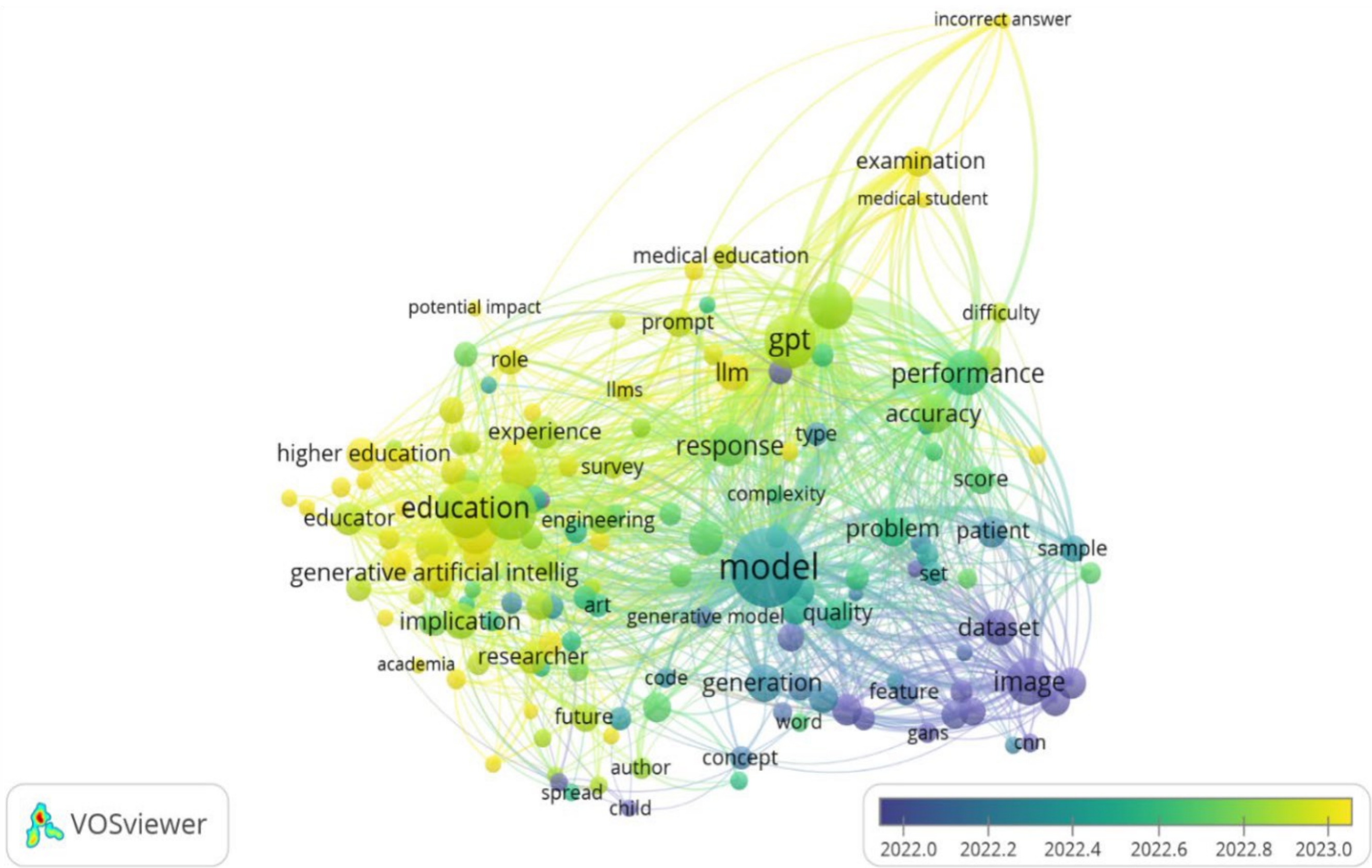

**Figure 9.** Map of co-occurrence (title and abstract).

The co-occurrence analysis (title and abstract) results yielded six clusters. The keywords grouped into clusters 3 (collaboration) and 6 (English, llms) are not informative. Cluster 2 consists of keywords such as algorithm, CNN, machine learning, and NLP, which suggest papers on algorithms and the underlying technologies that are being discussed extensively for the development of GenAI for intelligent educational technologies. This is crucial to integrate GenAI with virtual reality [61]. Cluster 4 is made up of keywords such as accuracy, answer, examination, incorrect answer, medical student, medical education, performance, and prompt, which are indicative of research conducted to examine the performance accuracy of GenAI systems on assessment/examination papers [62,63]. Cluster 5 consists of keywords such as experience, response, source, survey, and participants, which indicates survey research conducted so far that could be used to work towards an understanding of the perspectives of students/teachers on the use of GenAI [64]. Most notable is Cluster 1, which consists of keywords such as academia, academic integrity, critical thinking, and ethical consideration, which indicates publications related to the usage/implications of GenAI systems, linked to developing a greater understanding around academic integrity, the development of assessments, and better pedagogical practices in response to these emerging tools [10]. Furthermore, we deduce from Figure 9 that there is a movement in the themes from 2022 to 2023 (in yellow) towards incorrect answers, examination, medical students, medical education, LLMs, potential impact, higher education, and role. This indicates topic trends and, therefore, emerging research themes.

In summary, we use bibliometric indicators to quantitatively assess the publication patterns, research progress, and impact of academic literature. Using VOSviewer, we performed co-occurrence analysis, which was achieved using clustering. Clustering is the grouping of objects according to their similar attributes [65]. The authors [66] developed VOSviewer in 2010 [67] and demonstrated the use of VOSviewer to perform co-occurrence analysis of research publications using clustering. They adopted citation relationships as the similarity attribute to perform the cluster analysis. However, they acknowledged that the approach is limited when the period of analysis is relatively short. This is important to note considering that clustering analysis groups themes that are similar and does not explicitly uncover latent themes and, secondly, GenAI use in the HE context is relatively new. Thus, our study extends the co-occurrence analysis with TM to uncover latent themes. This agrees with the study by D'ascenzo et al. [68]. In subsequent sections, we applied the TM technique, namely LDA to complement the bibliometric analysis results, by distilling the current and future research pathways on GenAI for teaching and learning practice.

### 3.2. Topic Modelling Results

This section presents the LDA results. Beforehand, the researchers identified the optimal number of topics using a coherence score. To achieve this, a plot of the coherence scores (cv) against the number of topics (5–100) was produced, as shown in Figure A1 (in Appendix A). The plot shows that the model achieved optimality with a coherence score of 0.36 for 10 topics. Furthermore, we evaluated the model using coherence scores (Umass) and perplexity scores, which are reported in Table A1 (in Appendix A). Lastly, we used human interpretation, and researchers ascertained that the model performed best when the number of topics was set to 10. Therefore, Table 5 presents the topics in which the research documents were classed. Each of the topics contains 15 terms, which helped in profiling the 10 topics identified. Based on the terms and human interpretations, the topics were labelled. Furthermore, we produced a frequency plot, as shown in Figure 10, to understand the distribution of the research papers across the topics.

**Table 5.** TM results.

| Topic | Terms | Topic Label |
|---|---|---|
| 1 | Study, health, student, design, technology, control, medium, platform, tool, creation, issue, language, attention, practice, building. | Implications of GenAI |
| 2 | System, article, data, technology, language, study, interaction, risk, experience, scenario, application, management, approach, implication, challenge. | GenAI for education and research |
| 3 | Language, design, system, task, approach, study, result, method, process, assessment, development, domain, engineering, generation, framework. | Support system |
| 4 | Problem, tool, language, bias, student, practice, transformer, study, ability, llm, scenario, society, material, skill, level. | Bias and inclusion |
| 5 | Student, tool, study, educator, researcher, technology, language, data, challenge, concern, work, experience, knowledge, development, feedback. | Intelligent tutoring system |
| 6 | Data, machine, learning, analysis, result, development, method, application, work, area, technology, network, image, datasets, field. | Machine learning/AI application |
| 7 | The question, response, performance, answer, accuracy, gpt4, result, study, information, examination, knowledge, conclusion, case, chatbot, background. | Performance evaluation on exam questions |
| 8 | Technology, language, article, process, application, business, world, work, knowledge, course, information, capability, text, innovation, create. | GenAI for writing |
| 9 | Student, technology, practice, opportunity, question, university, healthcare, tool, article, challenge, concern, chatbots, response, impact, study. | Ethical and regulatory considerations |
| 10 | Image, network, method, application, study, generation, performance, data, result, accuracy, detection, approach, text, field, generate. | Deep learning/AI model |

This study interprets the following topics, as detailed below.

- Topic 1: Implications of GenAI (23 research papers)

The research documents grouped into topic 1 discussed several use cases of GenAI tools. For example, these research papers investigated the accuracy of content generated by GenAI, specifically for essay writing in health and computing disciplines. Furthermore, these papers investigated the effects of utilising GenAI in creating digital artifacts on students' understanding of AI literacy and their perception of social and ethical compliance. More specifically, a few papers emphasised the ethical implications, such as cheating. This topic made up 6.48% of the entire number of documents retrieved.

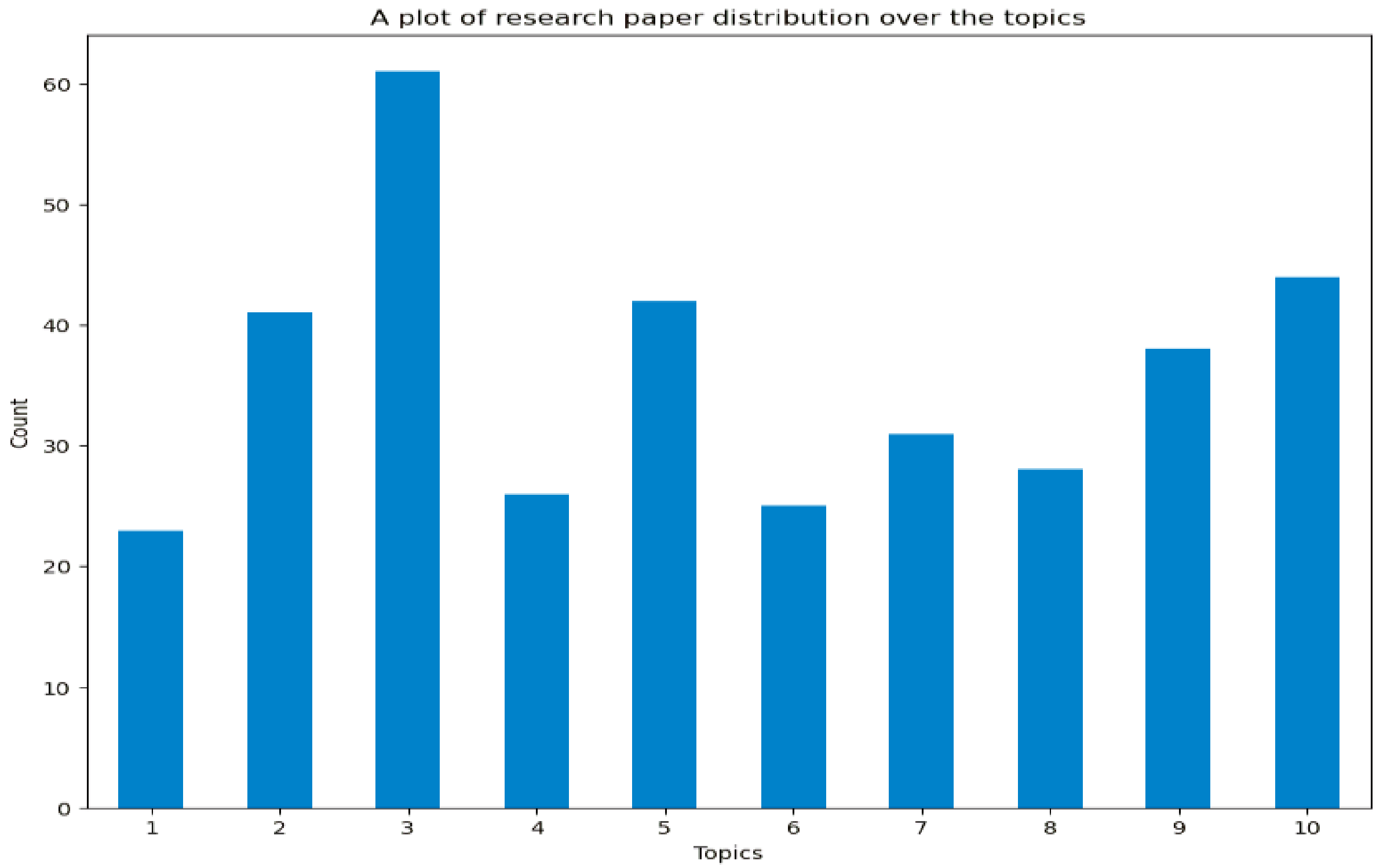

**Figure 10.** Topic distribution.

- Topic 2: GenAI for education and research (40 research papers)

The research papers grouped into topic 2 discussed the applications of GenAI to education and research in HE. These papers were specific to disciplines such as nursing, clinical science, ophthalmology, and radiation oncology. A few studies used GenAI systems to provide academic reviews of scientific papers. In addition, a few studies discussed policies and regulations on the adoption of AI tools. Specifically, we observed a study that proposed an AI ecological education policy framework for integrating GenAI tools into HE. Furthermore, these papers highlighted the potential benefits and drawbacks of integrating technology into education, providing insight into both the opportunities and challenges it presents. This topic made up 11.27% of the entire number of documents retrieved.

- Topic 3: Support system (60 research papers)

Topic 3 made up the highest proportion (16.91%) of research documents. The research papers in this group focused on the use of GenAI tools as a support system. For example, GenAI is a support system for education delivery, specifically for collaborative learning, exercise generation (to form a question bank), contract drafting, and administrative support. Other examples include the use of GenAI as a customer service (chatbot) system to support admissions to university. Furthermore, a few studies examined how GenAI systems may affect assessments across several disciplines, including medical and engineering education.

- Topic 4: Bias and inclusion (26 research papers)

The research documents categorised into this topic discussed how GenAI tools can be incorporated into education curricula, including teaching with GenAI across all levels of education. In addition, it was observed that the literature investigated teachers' and students' perspectives on inclusion. Furthermore, a few studies highlighted some setbacks related to GenAI tools. More specifically, bias and gender inequality were key themes discussed as a result of responses generated by GenAI tools. This topic made up 7.32% of the entire number of documents retrieved.

- Topic 5: Intelligent tutoring system (42 research papers)

Topic 5 research documents discussed the use of GenAI tools specific to teaching and learning practice. The research papers highlighted that GenAI tools can serve as a learning technology to support student learning outcomes. For example, students can engage the system to develop case studies including solutions on a particular subject and, thus, can critically reflect on the case studies. Further examples include using GenAI tools to provide a personalised learning experience, engagement, real-time interactivity, and feedback. These research papers emphasised the diverse applications, implications, and perspectives surrounding the integration of GenAI technologies into education, ranging from student modelling and feedback systems to language education research and beyond. This topic made up 11.84% of the entire number of documents retrieved and ranks third in terms of the highest proportion of documents.

- Topic 6: Machine learning/AI applications (25 research papers)

The research documents categorised in topic 6 presented the use of machine learning algorithms to develop AI applications. The papers investigated both the practical use and the challenges associated with GenAI technologies. For example, a study investigated how social work researchers can use such tools. Furthermore, another paper investigated the code generation performance of a system. This topic made up 7.05% of the entire number of documents retrieved.

- Topic 7: Performance evaluation on exam questions (30 research papers)

Research papers in this group examined the performance of GenAI tools for several examinations, specifically for medical education. Notable areas of medical subject matters being examined were plastic surgery graduate medical education examination, orthopaedic in-training examination questions, board-based questions on the Congress of Neurological Surgeon (CNS) self-assessment neurosurgery (SANS) exam, the American Board of Orthopedic Surgery examination, a fertility-related clinical prompt, the French language version of the European Board of Ophthalmology examination, Section 1 in the fellowship of the Royal College of Surgeons (trauma and orthopaedics) examination, the Peruvian national medical licensing examination, the Japanese medical licensing examination, the Japanese national examination for pharmacists, and the European Board of Ophthalmology (EBO) examination. Furthermore, a few studies experimented with GenAI systems as assistants to help in diagnosing and providing potential treatment suggestions for glaucoma and arthrosis. The papers also investigated whether there was agreement between GenAI and humans in terms of diagnosis and treatment suggestions. In addition, a few studies experimented with GenAI systems for teaching practice in the field of radiology. In total, the research documents in topic 7 amount to 8.46% of the entire number of documents retrieved.

- Topic 8: GenAI for writing (28 research papers)

Topic 8 consists of research documents that focus on opportunities for GenAI tools, specifically for academic and scientific writing. Some of the contexts discussed concern medical writing and drafting learning objectives. This topic made up 7.89% of the entire number of documents retrieved.

- Topic 9: Ethical and regulatory considerations (37 research papers)

The research documents in topic 9 discussed the challenges posed by GenAI tools in practice. The papers expressed the existing use of such tools in terms of timely responses/feedback and chatbots. However, they also discussed issues and challenges such tools can bring in regard to student learning. A few studies explored the ethical implications of using GenAI for teaching and learning and discussed pedagogical approaches to effectively integrate such tools, while ensuring ethical use and promoting meaningful learning outcomes. More specifically, the challenges discussed were academic integrity, hallucinations, plagiarism, misleading information, critical thinking, ethical issues, and data privacy. Many studies agreed that there is a pressing need to examine the ethical implications and establish appropriate regulations. The studies discussed the critical issues related to patient care, privacy, and professional ethics within the context of medical and healthcare education, making it relevant and important for discussions. This topic made up 10.42% of the entire number of documents retrieved.

- Topic 10: Deep learning/AI models (44 research papers)

The research documents categorised in topic 10 discussed the use of deep learning algorithms to develop AI applications. A popular example was the use of AI models for predicting academic performance. This topic made up 12.39% of the entire number of documents retrieved, which is the second highest topic discussed.

Overall, our TM results indicate that topics such as the implications of GenAI, GenAI for education and research, support systems, bias and inclusion, intelligent tutoring systems, machine learning/AI applications, performance evaluation on exam questions, GenAI for writing, ethical and regulatory considerations, and deep learning/AI models are the core themes of the research papers analysed. Our literature synthesis showed that a considerable number of studies reported the benefits of integrating GenAI tools to support personalised learning experiences, provide feedback, enhance student engagement, create learning activities, and support student learning outcomes [26]. The TM findings indicate that there are several proposed GenAI education policy frameworks for integrating these systems into HE. However, the issues of bias, gender inequality, misleading information, limitations to critical thinking, data privacy, plagiarism, and ethical and academic integrity are growing issues that are well stated in the literature [10].

## 4. Conclusions

This paper aims to provide an overview of the current state and progress in research on GenAI for teaching and learning in HE, through a systematic literature review. For this purpose, we used bibliometric indicators and TM to synthesise the literature. In response to RQ1, our findings show that more journal articles (72%) were published than conference papers (28%) in this genre. The results identified "Yogesh K. Dwivedi" as the author with the highest number of citations and the article "So what if ChatGPT wrote it? Multidisciplinary perspectives on opportunities, challenges, and implications of generative conversational AI" as the most cited research paper. Due to the emergence of GenAI tools, specifically the development of ChatGPT in November 2022 [5], our results evidenced the exponential growth in publications (in 2023) in the area of GenAI in HE. Moreover, the analysis showed that the Journal of Applied Learning and Teaching (JALT), the International Journal of Information Management (IJIM), and the New England Journal of Medicine (NEJM) were the most cited journals, with the United States of America, Australia, China, and the United Kingdom being the leading countries in terms of authorship. The latter is consistent with the findings in previous studies [55–60].

Aside from the current status of scholarly works on GenAI in learning and teaching in HE sectors that emerged from the types of publication (conference/journal), most cited authors/sources, and co-authorship (countries') perspectives, the findings also revealed the progression in the choice of keywords, based on the authors' keywords, as well as titles and abstracts. This invariably shed some light on the trend in the use and adoption of keywords, hence the direction of research pre- and post-GenAI (i.e., before 2022 when it was launched and afterwards).

In response to RQ2, the author's keyword analysis showed that academic integrity, assessment, chatbots, ChatGPT, education, ethics, generative pre-trained transformer, GPT-4, higher education, medical education, OpenAI, and prompt engineering are the topic trends. Similarly, the emerging themes identified in the title and abstract keyword analysis are incorrect answers, examination, medical student, medical education, LLMs, potential impact, role, and higher education. Furthermore, the TM results showed the core themes are the implications of GenAI, GenAI for education and research, support systems, bias and inclusion, intelligent tutoring systems, machine learning/AI applications, performance evaluation on exam questions, GenAI for writing, ethical and regulatory considerations, and deep learning/AI models.

Implications, Limitations, and Recommendations for Future Work

Theoretically, we evidenced that TM is a suitable technique to complement keyword co-occurrence analysis, because it provides an automated and efficient means of distilling latent themes. Furthermore, our approach demonstrates an alternative method to content/thematic analysis when reviewing the literature. In practice, our results benefit HE, stakeholders, and the research community to understand the current state of GenAI use. For example, the key topics identified, such as intelligent tutoring system, bias, and ethical considerations, are critical areas of focus. The substantial literature on GenAI in medical education indicates its potential use across various disciplines. Academics and students must understand GenAI's limitations to leverage its strengths effectively.

It is worth stating that this paper reviews literature in the English language and, thus, it might mean that important research in other languages was left out. This limitation means our study might have missed the non-English topic trends in GenAI research.

To enhance inclusivity, it is crucial to expand representation across journals, incorporating non-English publications. Longitudinal studies are needed to monitor evolving GenAI research trends continuously. Further exploration of the ethical implications and mitigation strategies is necessary to ensure responsible usage. Stakeholder engagement, particularly feedback from students and educators, is essential for refining GenAI tool development and aligning them with actual needs. Additionally, investigating the integration of GenAI modules into educational curricula and assessments, while considering the ethical aspects is vital for future research to ensure balanced knowledge and ethical usage. Overall, our results indicate that there is a significant amount of literature on the use of GenAI for medical education and, at the moment, this is disproportionate to other disciplines in HE. Therefore, the following recommendations are made:

- Furtherance to the keyword, title, and abstract analyses, significant studies that examine the performance of GenAI tools in medical and healthcare disciplines abound; such studies across disciplines are required/recommended in HE. With such multi-disciplinary and interdisciplinary studies, informed decisions on agreed guidelines towards the usage of GenAI systems in HE will emerge, hence the debate on GenAI will be well situated;
- This study revealed the countries with the largest number of publications, with none or low publications from developing countries. We, therefore, recommend that future publications be carried out in the area of GenAI through collaboration, especially in the global south;
- Academics need to understand the issues surrounding GenAI and develop strategies that will minimise its weaknesses but enhance its opportunities. Students also need to be aware of GenAI's limitations and shortcomings in terms of its non-ethical use and the implications on critical and analytical thinking, as well as the impairment of other soft skills. With this in mind, both tutors' and students' inputs will need to be successfully incorporated into GenAI tools for pedagogical practice;
- The development of LLM-based chatbots is growing. More recently, the development of Gemini has occurred, which is said to outperform ChatGPT-4 in most NLP tasks. This is yet to be ascertained in the HE domain. Thus, we recommend an experimental comparison of these GenAI tools for teaching and learning and assessment in terms of pedagogical practice;
- To successfully incorporate GenAI tools into teaching and learning practice, there is a need for users' input and perspectives with an interdisciplinary scope. Thus, there is a need for research synthesis from students' and academic tutors' perspectives to formulate the use of GenAI tools for teaching and learning pedagogical practice;

- Plagiarism detector systems like Turnitin have integrated AI content detectors into their system. However, the performance of such systems is not yet known. There is a need to examine the performance of Turnitin (and similar systems) to understand the extent to which these systems can identify AI-generated and human-written texts across several disciplines in HE;
- There is a need to update the curriculum in education [69]. However, there is a need to have a proper understanding of the potential impact of GenAI tools on the current curriculum. At this stage, it is not yet known whether including modules like an "Introduction to GenAI" in the curriculum will provide a balance between knowledge, usage, and ethics;
- To conclude, future research should be focused on interdisciplinary studies to develop guidelines for GenAI usage in HE. Experimental comparisons of advanced GenAI tools like Gemini and the performance of AI content detectors in plagiarism systems will be explored. Comparative studies should be conducted to assess the effectiveness of GenAI tools in educational settings, accurately. Updating curriculum and assessments to include GenAI topics, while assessing their impact on education, will be crucial for balanced knowledge and ethical usage.

**Author Contributions:** Conceptualisation, B.O. and K.I.Z.; methodology, B.O., K.I.Z., O.A., H.S. and O.O.; task creation /data curation, B.O., K.I.Z., O.A., H.S. and O.O.; moderation B.O. and K.I.Z.; writing—original draft preparation, B.O. and K.I.Z.; writing—review and editing B.O., H.S., K.I.Z., O.A. and O.O.; project management and supervision, B.O. All authors have read and agreed to the published version of the manuscript.

**Funding:** This research received no external funding.

**Institutional Review Board Statement:** Not applicable.

**Informed Consent Statement:** Not applicable.

**Data Availability Statement:** All the data are presented in the study.

**Conflicts of Interest:** The authors declare no conflicts of interest.

## Appendix A

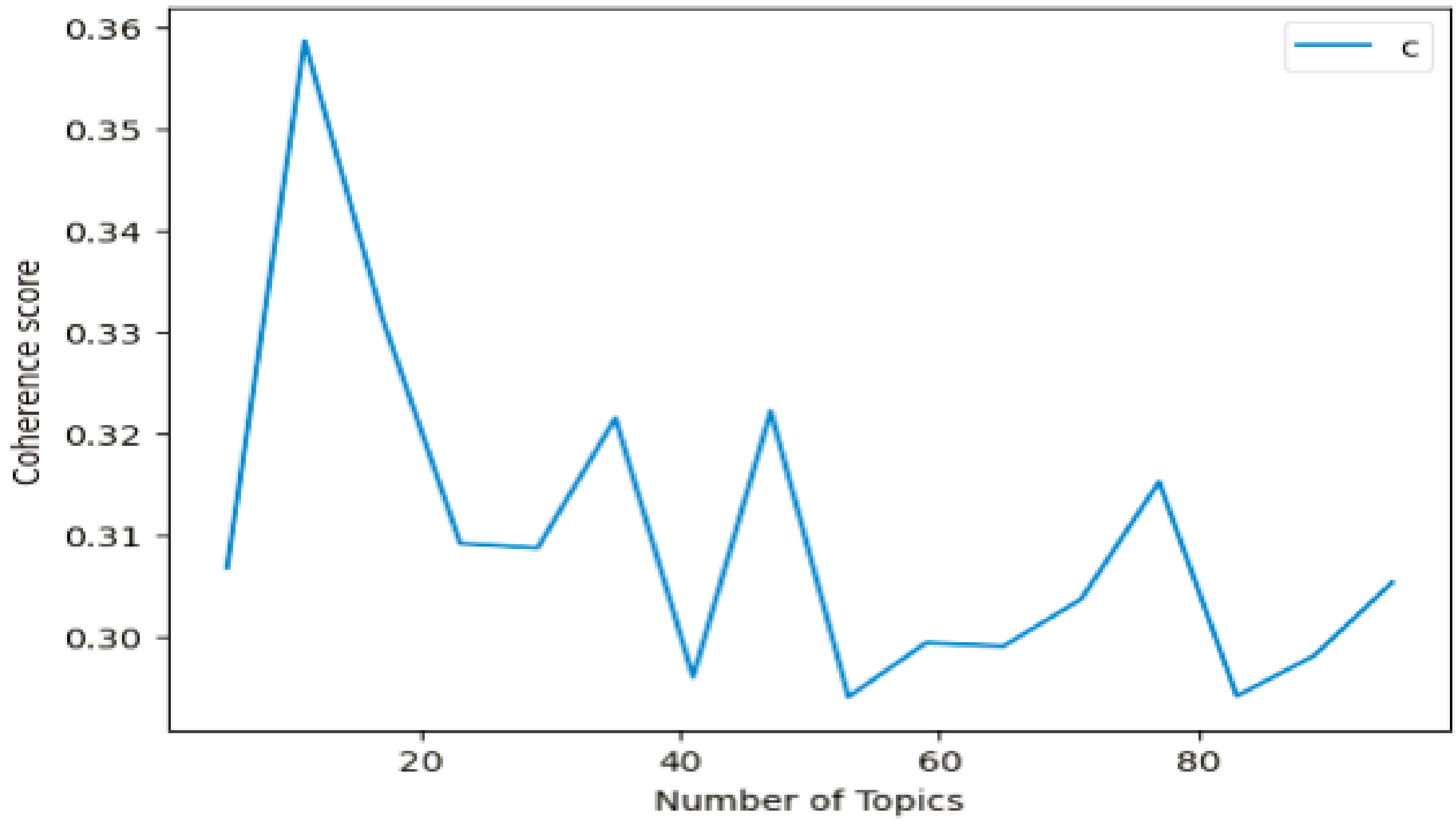

**Figure A1.** Plot of coherence score against the number of topics.

**Table A1.** Model evaluation.

| Metrics | Score |
|---|---|
| Coherence score (cv) | 0.3605 |
| Coherence score (Umass) | −2.0841 |
| Perplexity | −5.1338 |